\documentclass[10pt,conference]{IEEEtran}
\IEEEoverridecommandlockouts

\usepackage{cite}
\usepackage{amsmath,amssymb,amsfonts}
\usepackage{algorithmic}
\usepackage{graphicx}
\usepackage{textcomp}
\usepackage{xcolor}
\usepackage{booktabs} 
\usepackage[T1]{fontenc} 
\usepackage{multirow}
\usepackage{xspace}
\usepackage{color}
\usepackage{epstopdf}
\usepackage{threeparttable}
\usepackage{bm}
\usepackage{comment}
\usepackage{array}
\usepackage{tcolorbox}

\newcommand{\etal}{\hbox{\emph{et al.}}\xspace}
\newcommand{\eg}{\hbox{\emph{e.g.,}}\xspace}
\newcommand{\ie}{\hbox{\emph{i.e.,}}\xspace}

\newcommand{\appname}{CGAN4FL\xspace}

\usepackage{color}
\newboolean{showcomments}
\setboolean{showcomments}{true}
\ifthenelse{\boolean{showcomments}}
{ \newcommand{\mynote}[2]{
		\fbox{\bfseries\sffamily\scriptsize#1}
		{\small$\blacktriangleright$\textsf{\emph{#2}}$\blacktriangleleft$}}}
{ \newcommand{\mynote}[2]{}}

\def\BibTeX{{\rm B\kern-.05em{\sc i\kern-.025em b}\kern-.08em
    T\kern-.1667em\lower.7ex\hbox{E}\kern-.125emX}}
\begin{document}

\title{Mitigating the Effect of Class Imbalance in Fault Localization Using Context-aware Generative Adversarial Network}
\author{
\IEEEauthorblockN{Yan Lei$^{1,2\ast}$ \thanks{*Corresponding author.},
Tiantian Wen$^{1}$, Huan Xie$^{1}$, 
Lingfeng Fu$^{1}$, Chunyan Liu$^{1}$, Lei Xu$^{3}$, Hongxia Sun$^{4}$ }
\IEEEauthorblockA{$^1$School of Big Data \& Software Engineering Chongqing University, Chongqing, China
}
\IEEEauthorblockA{$^2$Peng Cheng Laboratory, ShenZhen, China}
\IEEEauthorblockA{$^3$Haier Smart Home Co., Ltd., Qingdao, China
}
\IEEEauthorblockA{$^4$Qingdao Haidacheng Purchasing Service Co., Ltd., Qingdao, China
}
 \IEEEauthorblockA{\{yanlei, tiantianwen, huanxie, lingfengfu, chunyanliu\}@cqu.edu.cn, \{xulei1, sunhongxia\}@haier.com}
}

\maketitle

\begin{abstract}
Fault localization (FL) analyzes the execution information of a test suite to pinpoint the root cause of a failure. 
The class imbalance of a test suite,
\ie the imbalanced class proportion between passing test cases (\ie majority class) and failing ones (\ie minority class), 
adversely affects FL effectiveness.

To mitigate the effect of class imbalance in FL,
we propose \appname: a data augmentation approach using \underline{C}ontext-aware \underline{G}enerative \underline{A}dversarial \underline{N}etwork for \underline{F}ault \underline{L}ocalization.
Specifically,
\appname uses program dependencies to construct a failure-inducing context showing how a failure is caused.
Then, \appname leverages a generative adversarial network to analyze the failure-inducing context and synthesize the minority class of test cases (\ie failing test cases).
Finally, \appname augments the synthesized data into original test cases to acquire a class-balanced dataset for FL.
Our experiments show that \appname significantly improves FL effectiveness,
\eg promoting MLP-FL by 200.00\%, 25.49\%, and 17.81\% under the Top-1, Top-5, and Top-10 respectively.

\end{abstract}


\section{Introduction}\label{intro}
To reduce debugging cost~\cite{planning2002economic,wong2016survey}, it is essential to develop effective approaches in the software debugging process.
In the literature, 
various fault
localization (FL) approaches (\eg~\cite{jones2004fault,jones2005empirical,li2019deepfl,li2021fault,naish2011model,sohn2017fluccs,wen2019historical,
wong2012software,zhang2019cnn}) have been proposed to pinpoint potential locations of faulty code over the past several decades.

\begin{figure}[htbp]\centerline{\includegraphics[width=0.5\textwidth]{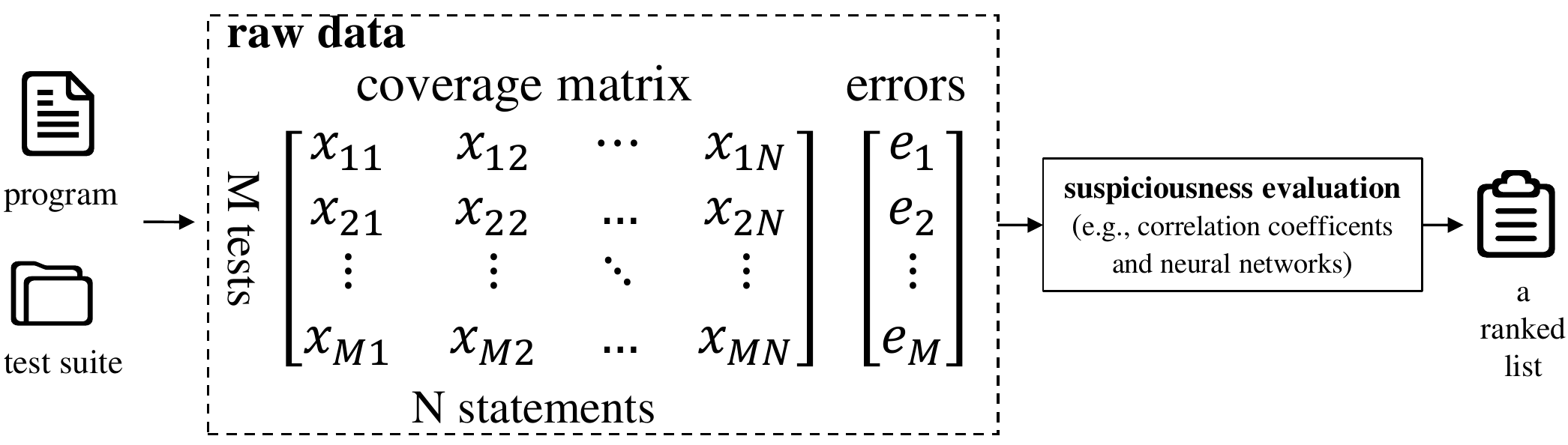}}
    \caption{Typical workflow of FL.}
    \label{fig1}
\end{figure}
\figurename{} \ref{fig1} shows the typical workflow of FL, \eg spectrum-based fault localization (SFL)~\cite{naish2011model,abreu2007accuracy} and deep learning-based fault localization (DLFL)~\cite{zhang2019cnn,zhang2021study,zheng2016fault}. 
FL executes the test cases of a test suite and collects the coverage information (denoted as coverage matrix) and test results (represented as errors) of each test case.  
The coverage matrix and errors are raw data for FL. 
In the raw data, 
each row of the coverage matrix represents the coverage information of a test case, 
and each column corresponds to the coverage information of a statement in all test cases of a test suite.
Specifically, for an element $x_{ij}$ in the coverage matrix,
$x_{ij}=1$ means that the \textit{i}-th test case executes the \textit{j}-th statement and otherwise $x_{ij}=0$;
for an element $e_i$ in the errors,
$e_i=1$ denotes that the \textit{i}-th test case is a failed test case and otherwise $e_i=0$.
After the raw data have been acquired, 
many FL approaches use them directly as input, and develop different suspiciousness evaluation algorithms (\eg SFL using correlation coefficients and DLFL using neural networks) to evaluate the suspiciousness value for each statement.
Finally, 
FL outputs a ranked list of program statements in descending order of suspiciousness values for manual or automated debugging~\cite{Xia2016Automated,kochhar2016practitioners,ghanbari2019practical,durieux2015automatic,martinez2016astor,le2012systematic,nguyen2013semfix,yi2017feasibility,tan2017codeflaws}.

Thus the raw data is indispensable for conducting effective FL. 
There are two classes of test cases: passing test cases and failing ones.
In practice, 
failing test cases are much less than passing test cases.
It leads to a class imbalance problem in the raw data,
\ie the data with failing labels are much less than the data with passing labels.
The existing studies~\cite{gong2012effects,zhang2017theoretical,zhang2021improving} have shown that the class imbalance problem inevitably introduces a bias~\cite{sun2009classification,he2009learning,chawla2009data} into the suspiciousness evaluation of FL and adversely affects FL effectiveness.

Since failing test cases are usually irregularly distributed and occupy a very small portion in the input domain, 
it is difficult to directly generate valid failing test cases in practice
to address the imbalance problem.
Inspired by the recent wide use of data augmentation approaches~\cite{goodfellow2020generative,antoniou2017data},
we can augment the raw data of FL by synthesizing new raw data with failing labels to acquire balanced raw data, \ie the class ratio between the raw data with failing labels and passing labels is balanced. 
In this way, 
a suspiciousness evaluation algorithm of FL can afterward utilize the class-balanced raw data to improve its accuracy.

Based on the above analysis,
we seek a data augmentation solution to the raw data of FL
for addressing the class imbalance problem in FL. 
Generative adversarial network (GAN)~\cite{goodfellow2020generative} is amongst the most popular data augmentation approach.
However, a failure-inducing context is useful for FL to acquire a reduced searching scope, 
and the original GAN does not consider a failure-inducing context into its data augmentation process, causing the augmentation to be potentially inaccurate.
It means that we should further incorporate a failure-inducing context into GAN to guide its data augmentation for FL.

Therefore, we propose \appname: a data augmentation approach using \underline{C}ontext-aware \underline{G}enerative \underline{A}dversarial \underline{N}etwork for \underline{F}ault \underline{L}ocalization, to mitigate the effect of class imbalance in FL.
\appname analyzes program dependencies via program slicing~\cite{agrawal1990dynamic} to construct a failure-inducing context, showing how a subset of statements propagates among each other to cause a program failure.
Then, \appname combines the failure-inducing context into a generative adversarial network to devise a context-aware generative adversarial network, 
which can synthesize the raw data of FL with failing labels.
Finally, we add the new synthesized failing raw data into the original raw data
to acquire class-balanced raw data,
where the data with failing labels have the same number 
as the data with passing labels.

To evaluate our approach \appname,
we design and conduct large-scale experiments on six representative subject programs. 
We apply \appname for six state-of-the-art FL approaches,
and further compare \appname with two representative data optimization approaches.
The experimental results show that our approach improves the effectiveness of the six state-of-the-art FL approaches, and outperforms the two representative data optimization approaches.
Specifically, the experimental results indicate that our approach compared with the six state-of-the-art FL approaches improves the effectiveness of fault localization by 125.34$\%$, 56.32$\%$, and 59.71$\%$ respectively on Top-1, Top-5, and Top-10 metrics on average.

The main contributions of this paper can be summarized as follows:
	\begin{itemize}
	\item {We propose a data augmentation approach \appname, which synthesizes failing raw data for mitigating the effect of class imbalance problem in FL.}
		
	\item  {We present a context-aware generative adversarial network which integrates a failure-inducing context into the data augmentation process of a generative adversarial network to guide data synthesization for FL.}
		
	\item {We conduct large-scale experiments and compare our approach with six state-of-the-art FL approaches and two representative data augmentation approaches, 
	showing that \appname significantly improves fault localization effectiveness.}
    \item {We open source the replication package online including all relevant code\footnote{https://anonymous.4open.science/r/CGAN4FL-B448}.}
	\end{itemize}
	
	The remainder of this paper
	is organized as follows.
	Section~\ref{background} introduces background information.
	Section~\ref{approach} presents our approach \appname.
	Section~\ref{experiments} and Section~\ref{discussion} show the experimental
	results and discussion.
	Section~\ref{relatedwork} discusses related work
	and Section~\ref{conclusion} draws the conclusion.
\section{Background}\label{background}
\subsection{Generative Adversarial Network}
\begin{figure}[htbp]\centerline{\includegraphics[scale=0.4]{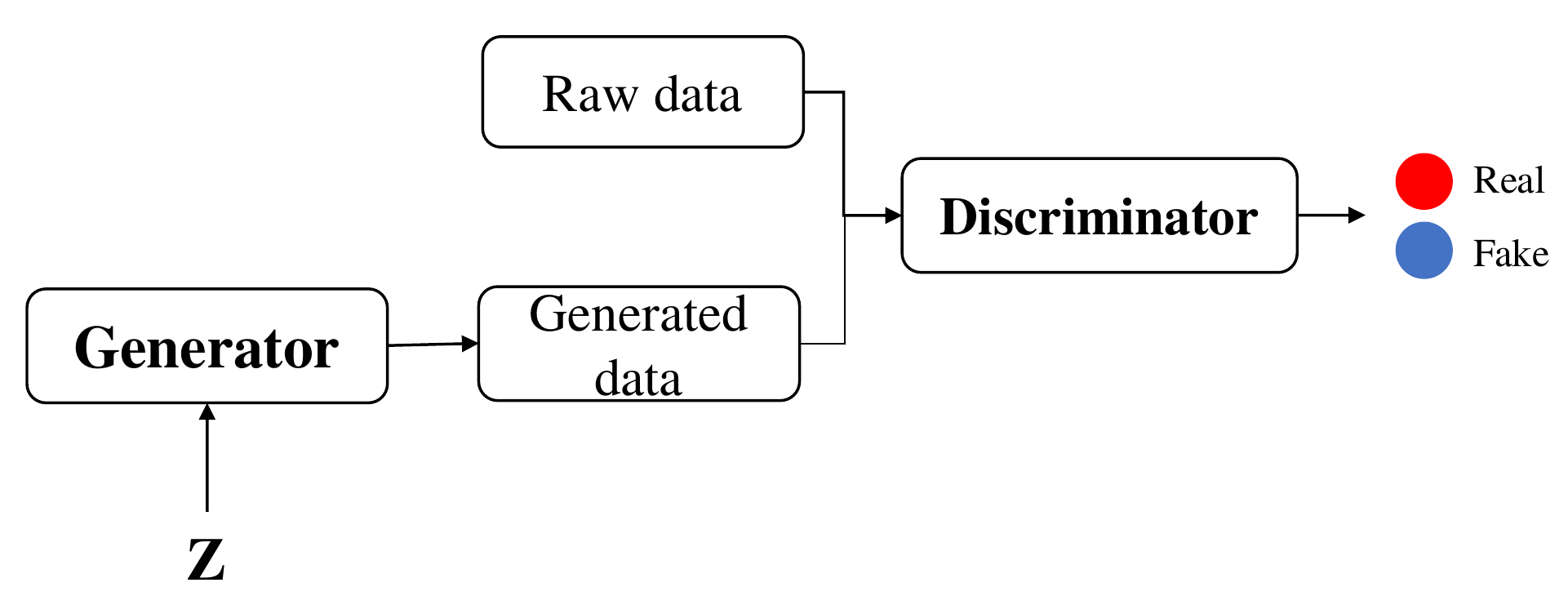}}
    \caption{Basic framework of a GAN.}
    \label{fig2}
\end{figure}
Generative Adversarial Network (GAN) \cite{goodfellow2020generative} is a deep learning framework that learns to generate adversarial data. 
\figurename{}~\ref{fig2} shows the basic framework of a GAN. 
GAN contains two components: the generator $G$ and the discriminator $D$. 
The generator $G$ is responsible for generating fake data that look like real data from the latent variable $z$ while the discriminator $D$ distinguishes whether the data belongs to raw data or generated data as accurately as possible. 
$G$ and $D$ are aggressive since they compete in order to accomplish their own objectives.
The purpose of the model training is to minimize the loss of $G$ and maximize the loss of $D$. 
Specifically when a generator has a lower loss, it means that the generated data is almost identical to real data; and when a discriminator obtains a higher loss, it means that it is hard to discriminate between real and generated data. 

This adversarial learning situation can be formulated as Eq.~(\ref{formula1}) with parametrized networks $G$ and $D$.
\begin{equation} 
    \begin{split}
        \min _{G} \max _{D} V(G,\,D)= \mathbb{E}_{x \sim p_{\text {data }}}[\log D(x)]+\\
        \mathbb{E}_{z \sim p_{z}}[\log (1-D(G(z))]
    \end{split}
    \label{formula1}
\end{equation}

In Eq.~(\ref{formula1}), $p_{data} \left (x  \right ) $ and $p_{z} \left ( z \right ) $ represent the real data probability distribution defined in data space $\mathcal{X} $ and the probability distribution of $z$ defined in latent space  $\mathcal{Z} $.
$V\left ( G,\, D \right ) $ is a binary cross entropy function that is commonly used in binary classification problems\cite{mao2017least}. 
It should be noted that $G$ maps $z$ from $\mathcal{Z} $ into the element of $\mathcal{X} $, while $D$ takes an input $x$ and determines whether $x$ is real data or fake data generated by $G$.

Since the goal of $D$ is to identify real or fake samples, $V\left ( G,\,D \right ) $ is a natural choice for this goal with its ability to solve binary classification problems. From the perspective of $D$, if the input data is real, the output of $D$ should be close to maximum; 
if the input data comes from $G$, $D$ will minimize its output. Thus, the $\log_{}{\left (   1-D\left ( G\left ( z \right )  \right )\right ) } $ term is added to Eq.~(\ref{formula1}). 
At the same time, $G$ plans to cheat $D$ and thus it tries to maximize $D$’s output when the input data is generated by it. Consequently, $D$ tries to maximize $V\left ( G,\, D \right ) $  while $G$ tries to minimize $V\left ( G,\, D \right ) $, forming a type of adversarial relationship.

The existing studies~\cite{gong2012effects,zhang2017theoretical,zhang2021improving} have shown that the class imbalance problem of raw data adversely affects FL effectiveness, and it is crucial to address the class imbalance problem in FL.
One of the best merits of GAN is that they generate data that is similar to real data. Due to this merit, 
they have many different applications in the real world,
\eg generating images, text, audio, and video that are indistinguishable from real data\cite{antoniou2017data,balaji2019conditional,zhang2016generating,kim2021fre}.
Inspired by the merit of GAN,
our study utilizes the ability of GAN to mitigate the effect of class imbalance in FL
via synthesizing minority class data for acquiring a class-balanced dataset.

\subsection{Fault Localization}\label{backgroundFL}
Fault localization (FL) typically collects and abstracts the runtime information of a test suite as the raw data (\ie the coverage matrix and errors in \figurename{}~\ref{fig1});
then takes the raw data as input to evaluate the suspiciousness value for each statement; finally outputs a ranked list of program elements in descending order of suspiciousness values. 
There are many granularity types of program elements, 
\eg statements, methods, and files.
Our study adopts the most widely-used granularity type
of program elements, \ie statements.
This section will introduce two popular FL techniques (\ie spectrum-based fault localization and deep learning-based fault localization), and they all use the raw data in \figurename{} \ref{fig1} as input for suspiciousness evaluation. 
Our experiments will also apply our approach \appname for these FL techniques to evaluate its effectiveness.

\textbf{Spectrum-based Fault Localization (SFL).} SFL~\cite{naish2011model,abreu2007accuracy} has been intensively studied in the literature. 
The basic idea of SFL is that the  suspiciousness of a statement should increase when it is executed more frequently by failing test cases; its suspiciousness should decrease when it is executed more frequently by passing test cases. 

To implement the above idea, 
SFL uses the raw data (\ie coverage matrix and errors) to define the four variables for each statement.
Let $s_j$ be a statement in the program.
Eq.~(\ref{eqSFLVars}) defines the four variables for $s_j$ as follows:

\begin{equation}\label{eqSFLVars}
\begin{aligned}
   a_{np}(s_j) &= | \{i| x_{ij} = 0 \wedge e_{i} = 0\}|  \\
   a_{nf}(s_j) &= | \{i| x_{ij} = 0 \wedge e_{i} = 1\}|   \\
   a_{ep}(s_j) &= | \{i| x_{ij} = 1 \wedge e_{i} = 0\}|  \\
   a_{ef}(s_j) &= | \{i| x_{ij} = 1 \wedge e_{i} = 1\}|
\end{aligned}
\end{equation}

Where, $a_{np}(s_j)$ and $a_{nf}(s_j)$
represent the numbers of test cases
that do \emph{not} execute the statement $s_j$
and return the \emph{passing} and \emph{failing} test results,
respectively;
$a_{ep}(s_j)$ and $a_{ef}(s_j)$ stand for the numbers of test cases
that \emph{execute} $s_j$,
and return the \emph{passing} and \emph{failing} testing results, respectively.

Based on the four variables,
SFL devises many suspiciousness evaluation formulas to evaluate the suspiciousness of a statement\cite{naish2011model,hao2005similarity,hao2005eliminating,xie2013theoretical,yoo2012evolving}.
The existing work~\cite{pearson2017evaluating} has empirically identified the three most effective SFL formulas (\ie Ochiai\cite{abreu2007accuracy}, DStar\footnote{The ‘*’ in Dstar formula is usually assigned to 2.}\cite{wong2013dstar}, and Barinel\cite{abreu2009spectrum}) in locating real faults.
Since our study focuses on locating real faults, our experiments use the three SFL formulas.
Based on the four variables defined in Eq.~(\ref{eqSFLVars}), Eq.~(\ref{ochiai}), Eq.~(\ref{dstar}) and Eq.~(\ref{barinel}) show the definitions of the three SFL formulas to compute the suspiciousness of a statement $s_j$.

\begin{equation}\label{ochiai}
    Ochiai(s_j) =\frac{a_{ef}(s_j)}{\sqrt{\left ( a_{ef}(s_j)+a_{nf}(s_j)  \right ) \times\left ( a_{ef}(s_j)+a_{ep}(s_j) \right )  } } 
\end{equation}
\begin{equation}\label{dstar}
    Dstar (s_j)= \frac{a_{ef}(s_j)^{*} }{a_{ep}(s_j)+ a_{nf}(s_j)}
\end{equation}
\begin{equation}\label{barinel}
    Barinel (s_j)= 1-\frac{a_{ep}(s_j)}{a_{ep}(s_j)+a_{ef}(s_j)}
\end{equation}

\begin{figure*}[htbp]
    \centerline{\includegraphics[scale=0.45]{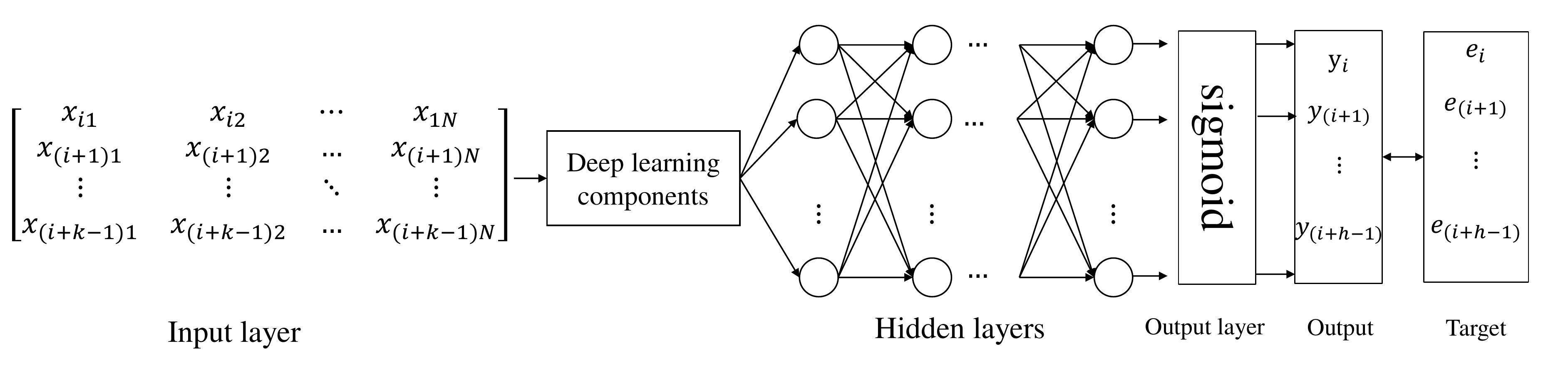}}
    \caption{Architecture of DLFL.}
    \label{fig3}
\end{figure*}
\textbf{Deep Learning-based Fault Localization (DLFL).} 
DLFL~\cite{zhang2019cnn,zhang2021study,zheng2016fault} has recently attracted much attention and acquired promising results. 
The basic idea of DLFL is that it utilizes the learning ability~\cite{guo2017semantically,hinton2006reducing} of neural networks to learn a FL model
which reflects the relationship between a statement and a failure. 
\figurename{}~\ref{fig3} shows the typical architecture of DLFL. DLFL usually has three parts: 
input layer, deep learning component with several
hidden layers, and output layer. Next, we will introduce three representative DLFL approaches, \ie MLP-FL\cite{zheng2016fault}, CNN-FL\cite{zhang2019cnn} and RNN-FL\cite{zhang2021study}, and our experiments apply our approach \appname to the three DLFL approaches to evaluate its effectiveness.

In the input layer, DLFL takes the raw data (\ie coverage matrix and errors) in \figurename{} \ref{fig1} as input. 
Specifically, \emph{k} rows of the coverage matrix and its corresponding errors vector, \ie the coverage information of \emph{k} test cases and their corresponding test results, are used as input. As shown in \figurename{} \ref{fig3}, these \emph{k} test cases are the rows starting from the \textit{i}-th row, where $i\in \left \{ 1,1+  k,1+2k,\dots ,1+\left ( \left \lceil M/k  \right  \rceil -1 \right )\times k  \right \} $. In the part of deep learning components, different fault localization methods use different neural networks.
For example, MLP-FL\cite{zheng2016fault} uses multi-layer perceptron, CNN-FL\cite{zhang2019cnn} adopts convolutional neural network, and RNN-FL\cite{zhang2021study} utilizes recurrent neural network. In the output layer, the model uses \emph{sigmoid} function to make sure that the output results are between 0 and 1.
Each element in the result of \emph{sigmoid} function could differ from its corresponding element in the target vector. 
The parameters of the model are updated using the backpropagation algorithm with the intention of minimizing the difference between training result $y$ and errors vector $e$ (\ie the errors in \figurename{}~\ref{fig1}). The network is trained iteratively. Finally, DLFL learns a trained
model, which can reflect the relationship between a statement and a failure. 
With the trained model, DLFL can evaluate the suspiciousness value for statements.

\section{Approach}\label{approach}
\begin{figure*}[htbp]
\centerline{\includegraphics[scale=0.5]{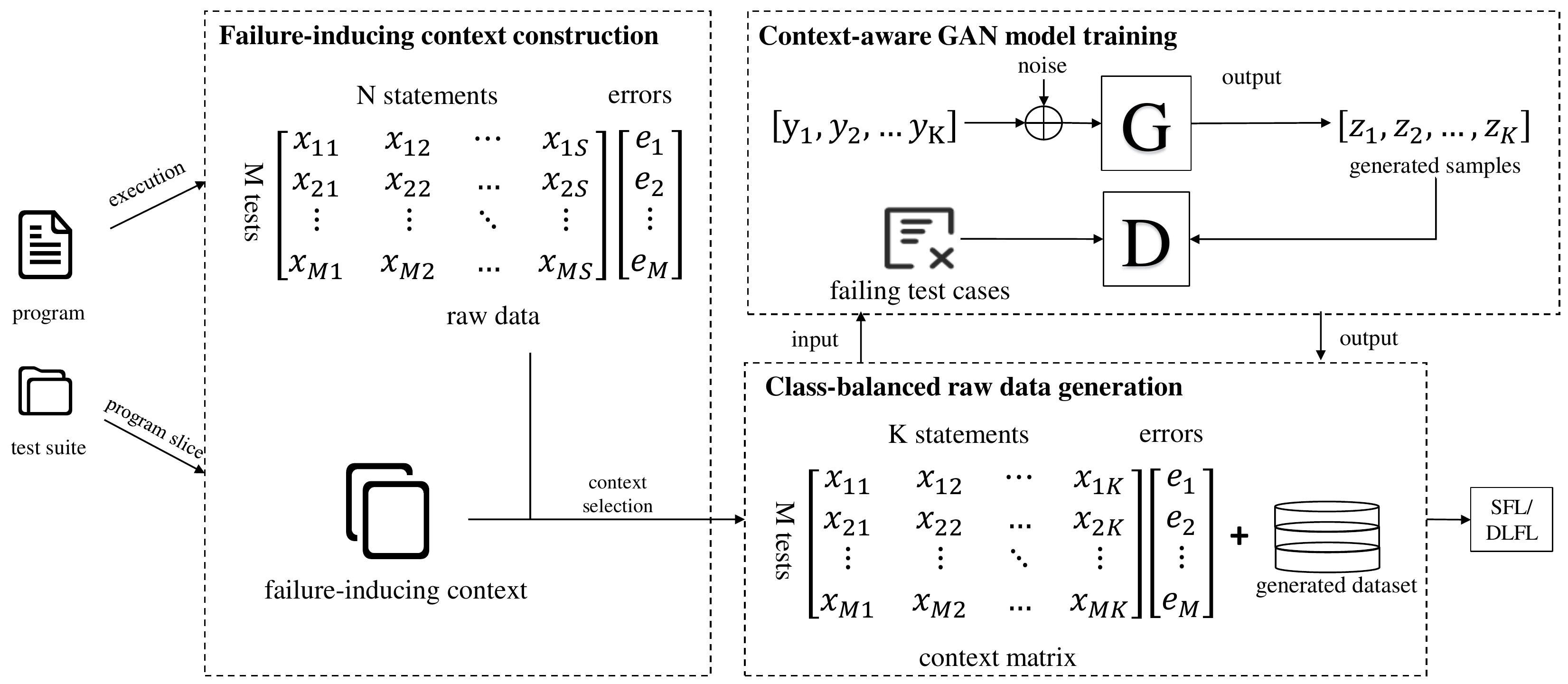}}
    \caption{Architecture of \appname.}
    \label{fig4}
\end{figure*}
This section will introduce our approach \appname: 
a data augmentation approach using Context-aware Generative Adversarial Network for Fault Localization.
As shown in \figurename{}~\ref{fig4},
\appname first uses program dependencies to construct a failure-inducing context;
then combines the failure-inducing context into the GAN training to learn a context-aware GAN model;
finally uses the trained context-aware GAN model to generate new failing raw data until a class-balanced dataset is acquired,
where the raw data with failing labels have the same number as the raw data with passing labels.

\subsection{Failure-inducing Context Construction}
A failure-inducing context shows 
how a subset of program elements (\eg statements) act on each other to cause a failure,
and it is useful for FL to acquire a reduced searching scope.
Therefore,
{\color{blue} we} intend to integrate a failure-inducing context into the GAN training to guide its data augmentation.
To implement the above idea,
\appname adopts the widely-used program dependencies via program slicing~\cite{xu2005brief} 
to construct a failure-inducing context,
showing the dependencies between a subset of statements that cause a failure.
The program slicing technique~\cite{xu2005brief} extracts the program dependencies among statements 
to pick out a subset of statements
whose execution leads to the incorrect output (\ie a failure). A few approaches have evaluated the effectiveness of dynamic slices in fault localization\cite{zhang2005experimental,zhang2007study,li2020more}.
The subset of statements is a program slice, \ie a failure-inducing context in our approach \appname. Thus, we define a failure-inducing context as follows:
\noindent
\begin{center}
	\fbox{
		\parbox{0.46\textwidth}{
			\textbf{A failure-inducing context:}
			statements that directly or indirectly affect the computation of the faulty output value of a failure through chains of dynamic data and/or control dependencies.
	}}
\end{center}

To compute a failure-inducing context using program slicing,
we use the following slicing criterion \textit{contextSC}.
\begin{equation}
    contextSC = \left ( outStm,outVar,failTest \right ) 
    \label{formula6}
\end{equation}

In Eq.~(\ref{formula6}), $outStm$ 
is an output statement whose value of a variable (\ie $outVar$) is incorrect in the execution of a failing test case (\ie $failTest$).
Dynamic slicing collects runtime information along the execution path of a test case, \ie
the set of executed statements of a test case. 
It means that a test case with a smaller set of executed statements is
usually easier for a dynamic slicing tool to perform efficient instrumentation and produce compressed traces for space
optimization. Thus, for multiple failing test cases, the one with the least executed statements usually is beneficial for
the efficiency of constructing a failure-inducing context. From the efficiency aspect, \appname will choose the failing test case having the least executed statements to construct a slicing criterion in Eq.~(\ref{formula6}).

Suppose that a failure-inducing context has $K$ statements. 
It means that these $K$ statements interact with each other to cause an incorrect output (\ie a program failure). 
Since the statements not in the failure-inducing context do not affect the incorrect output, 
we combine the failure-inducing context 
via keeping the coverage information of these statements in the failure-inducing context and removing others.
Thus, \appname finally acquires a new $M \times K$ matrix called context matrix
which records the execution information of the failure-inducing context in
the test suite. 

\subsection{Context-aware GAN Model Training}
After constructing a failure-inducing context,
we acquire a $M \times K$ context matrix
from the original $M \times N$ matrix (\ie original raw data).
The context matrix shows the runtime information of these statements whose
execution leads to the incorrect output of a program. 
\appname uses the context matrix as the input of the GAN model,
\ie \appname trains the GAN
to generate a new synthesized vector
by the discriminating network $D$ and the generating network $G$ with all failing test cases selected 
from the context matrix as samples. It means that 
\appname will learn the features of all failing test cases,
\ie the newly synthesized test cases will cover the common feature of all failing test cases. 
Thus, 
\appname will mark the newly synthesized test cases (\ie the newly synthesized vectors with the same structure of raw data) as failing labels.
We add the new failing synthesized test cases to the context matrix and 
form a new matrix (\ie new raw data) whose failing data and passing data are balanced. 
Finally, the new raw data are used as the new input for the FL approach (\eg SFL and DLFL) to improve its effectiveness.

The model training part of \figurename{}~\ref{fig4} shows the specific \appname training procedure. \appname trains $D$ (discriminator) first to initiate the training procedure of the GAN model. 
\appname selects the \textit{K}-dimensional vector $\left [ y_{1},\,y_{2},\,\dots,\,y_{K}  \right ] $, and inputs it into $G$ (generator) after noise processing to obtain the generated data $\left [z_{1},\,z_{2},\,\dots,\,z_{K}  \right ] $ (\ie the synthesized data). 
Then, 
it further selects the original failing test cases in the original raw data (\ie $\left[x_{i1},\, x_{i2},\, \dots , \,x_{iK} \right ] $, $i\in\left \{ 1,\, 2,\, \dots ,\, M \right \} $ and $e_{i}=1$) as the real data, and uses the generated data $\left [ z_{1}, \,z_{2},\, \dots , \,z_{K}  \right ] $ as the generated sample to be spliced together and input them into $D$ (discriminator). 
$D$ gives label 1 to real data $\left [ x_{i1},\,x_{i2},\,\dots ,\,x_{iK} \right ] $, and 0 to generated data $\left [ z_{1},\,z_{2},\,\dots ,\,z_{K} \right ] $. The difference between $D$'s output score and the label is trained using loss backpropagation.
After $D$ training is complete, 
\appname starts $G$ training 
with the fixed parameters of $D$. 
In the $G$ training process, $D$ and $G$ are regarded as a whole. $\left [ y_{1},\,y_{2},\,\dots ,\,y_{K}  \right ] $  processed by the noise is used as the input, and then $G$ outputs generated data $\left [ z_{1},\,z_{2},\,\dots ,\,z_{K}    \right ] $. The discriminator $D$ with fixed parameters is used for scoring. The difference between the output score and label 1 is used as the loss backpropagation to train $G$.

Throughout the GAN model training process, 
$G$ is weak at the beginning, and $D$ can easily distinguish between real data and generated data. With the gradual increase of training $G$, $D$ cannot distinguish between real data and generated data. Eq.~(\ref{formula1}) is the training process of the minimax two-player game between generator $G$ and discriminator $D$. $D$ maximizes the objective function to identify whether the generated data  $\left [ z_{1},\, z_{2},\, \dots, \,z_{K}    \right ] $ are fake. In contrast, $G$ constantly minimizes the distribution difference between the real data and the generated data, \ie minimizing $D$'s discrimination of generated data. Finally, a Nash equilibrium\cite{ratliff2013characterization} is reached.
\subsection{Class-balanced Raw Data Generation}
After context-aware GAN training,
\appname learns a context-aware GAN model which generates synthesized failing data (\ie the new synthesized failing vectors with the same structure as the original raw data) for FL.
The trained model will generate synthesized failing data and add them to the original raw data until we acquire a class-balanced dataset,
where the number of failing vectors and 
the number of passing vectors are the same 
in the new raw data.
\appname inputs the new class-balanced raw data into the suspiciousness evaluation algorithm of the FL approach to mitigate the effect of the class imbalance problem in FL.
Finally, 
with the new class-balanced raw data,
FL outputs a ranked list of all statements in descending order of suspiciousness values.
\subsection{An Illustrative Example}
\begin{figure*}[htbp]
    \centerline{\includegraphics[width=0.8\textwidth]{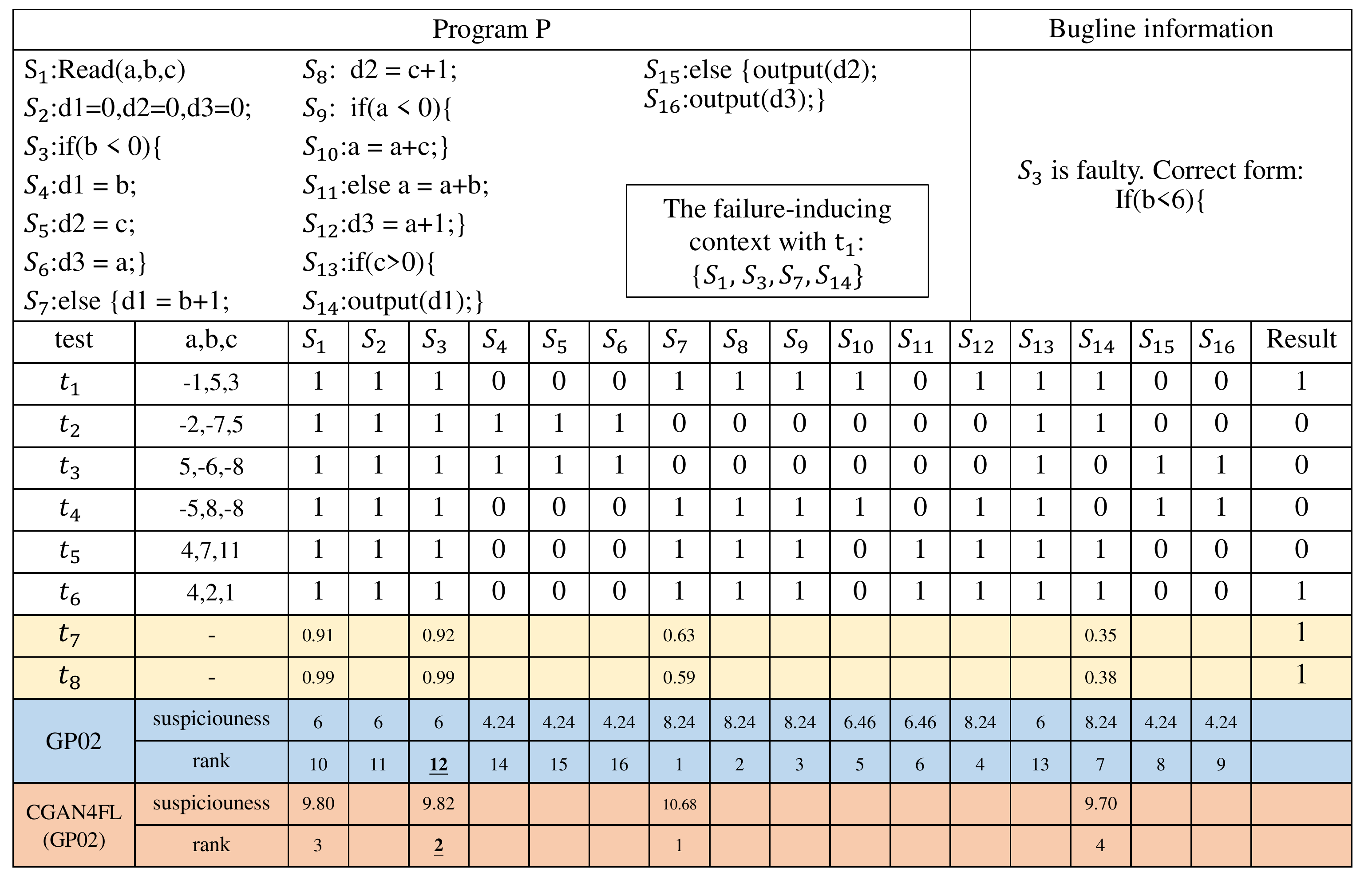}}
    \caption{An example illustrating \appname.}
    \label{fig5}
\end{figure*}
To illustrate how the methodology of \appname works, 
\figurename{} \ref{fig5} shows an example of applying \appname.
As shown in \figurename{} \ref{fig5}, 
there is a faulty program \textit{P} with 16 statements including a fault at line 3, in which the number 0 should be 6 instead. We use one SFL approach (\ie GP02\cite{xie2013theoretical}) to locate the faulty statement for our illustrative example.  
The cells below each statement indicate whether the statement 
is executed by the test case or not (\ie 0 for not executed and 1 for executed). The cells below the `Result' which is the
errors vector for the coverage matrix as shown in \figurename{} \ref{fig1} represent whether the test result of a test case is failing or passing (\ie 1 for 
falling and 0 for passing).
The original test suite is class-imbalanced since it has four passing test cases (\ie $t_2$, $t_3$, $t_4$, and $t_5$) and two failing test cases (\ie $t_1$ and $t_6$). 

For acquiring a class-balanced dataset,
we need to generate two pieces of new raw data with failing labels. 
\appname first uses the failing test case $t_1$ to compute the failure-inducing context using program slicing.
According to Eq.~(\ref{formula6}), 
we set ($S_{14}$,\,d1,\,$t_1$) as the slicing criterion since the output value of the variable d1 in the output statement $S_{14}$ is incorrect when executing the failing test case $t_1$.  
As shown in \figurename{} \ref{fig5}, the failure-inducing
context regarding $t_1$ is $\left \{ S_{1},\,S_{3},\,S_{7}\,,S_{14}\right \}$. Then, based on the failure-inducing context, \appname uses GAN to generate two pieces of synthesized failing raw data (\ie $t_7$ and $t_8$) marked with yellow in \figurename{} \ref{fig5}.
Finally, \appname adds the two failing synthesized failing test cases into the context matrix to form a new raw data, 
and GP02 uses the new raw data to conduct the suspiciousness evaluation for each statement. 

The bottom rows are the FL results of original GP02 and GP02 with \appname,
\ie the two ranked lists of statements in descending order of suspiciousness in \figurename{} \ref{fig5} marked with different colors.
Without using \appname,
the ranked list of the statements using GP02 marked with blue is $\{ S_{7},S_{8},S_{9}, S_{12},S_{10},S_{11},S_{14},S_{15},S_{16},S_{1},S_{2},S_{3},S_{13},S_{4},$ $S_{5},S_{6}\}$.
After applying our approach \appname, 
the ranked list of the statements using GP02 is $\left \{ S_{7}, S_{3},S_{1},S_{14}\right \}$. 
We can observe that the faulty statement $S_{3}$ is ranked 12th place with the original raw data while \appname ranks the faulty statement $S_{3}$ 2nd place. 
It means that \appname
yields better FL results than the original GP02, 
mitigating the effect of the class imbalance in FL.

\section{Experiments}\label{experiments}
\subsection{Datasets}
To evaluate the effectiveness of our approach \appname, we adopt the Defects4J~\cite{just2014defects4j}
that has been widely used in the software testing and debugging community~\cite{b2016learning,martinez2016astor,xie2022universal,zhang2017boosting}. We use all the six representative subject programs of Defects4J\footnote{https://github.com/rjust/defects4j} (
\ie Chart, Closure, Math, Mockito, Lang, and Time) and all faults from these programs are real faults. 

\tablename~\ref{table1} summarizes the information of the six subject programs.
For each program, it lists a brief functional description (column `Description’), the number
of faulty versions used (column `Versions’), the number of
thousand lines of statements (column `LoC(K)’), the number
of test cases (column `Test’).
Since it is time-consuming to 
collect the inputs (\ie raw data) of the six large programs of Defects4J, 
we reuse the coverage matrix and the errors collected by Pearson~\etal~\cite{pearson2016evaluating}.

\subsection{Experiment Settings}
For every faulty version of the six subject programs shown in \tablename~\ref{table1}, we set the training period of context-aware GAN model to 1,000, and the dimension of hidden variable $z$ defined in latent space $\mathcal{Z} $ to 100. 
Our experiments were conducted on a 64-bit Linux server with 40 cores of
2.4GHz CPU and 252GB RAM. The operating system is Ubuntu 20.04.
\subsection{Evaluation Metrics}
We adopt the four widely-used FL evaluation metrics to evaluate the effectiveness of our approach. 
Their definitions are as follows:
\begin{itemize}
    \item Number of Top-K\cite{heris2019effectiveness,kuccuk2021improving}: It is the number of faulty versions with at
     least one faulty statement that is within the first K position of the rank list produced by the FL technique. In the previous study, 
     many respondents view fault localization as successful only if it can localize bugs in the top 10 positions from
      a practical perspective \cite{heris2019effectiveness,kuccuk2021improving}. Following the prior work~\cite{heris2019effectiveness,kuccuk2021improving}, we assign K with the value of 1, 5, and 10 for our evaluation. A higher value of Top-K means better FL effectiveness.
\begin{table}[!ht]
    \centering
    \renewcommand\arraystretch{1.5}
    \caption{Subject programs}
    \resizebox{0.49\textwidth}{!}{
        \begin{tabular}{c|c|c|c|c}
            \hline
                Program & Description & Versions & LoC(K) & Test \\ \hline
                Chart & Java chart library & 26 & 96 & 2205 \\ \hline
                Lang & Apache commons-lang & 65 & 22 & 2245 \\ \hline
                Math & Apache commons-math & 106 & 85 & 3602 \\ \hline
                Closure & Closure compiler & 133 & 90 & 7927 \\ \hline
                Time & Standard date and time library & 27 & 28 & 4130 \\ \hline
                Mokito & Mocking framework for Java & 38 & 67 & 1075 \\ \hline
                Total & - & 395 & 388 & 21184 \\ \hline
        \end{tabular}
    }
    \label{table1}
\end{table}
    \item Mean Average Rank (MAR)\cite{li2019deepfl}: For a faulty version, the average rank is the mean rank of all faulty statements in the rank list. 
    MAR is the mean average value for the project that includes several faulty versions. A lower value of MAR indicates better FL effectiveness.

    \item Mean First Rank (MFR)\cite{li2019deepfl}: It first computes the rank that any of the statements are located first for a faulty version.
     Then compute the mean value of the ranks for the project. 
     A lower value of MFR shows better FL effectiveness.
     
     \item Relative Improvement (RImp)\cite{zhang2019cnn}: This metric can see the improvement of one fault localization approach relative to another fault localization approach. It is to compare the total number of statements that need to be examined to find the first faulty statement using \appname versus the number that needs to be examined by using baselines. 
    A lower value of RImp indicates better FL effectiveness.
\end{itemize}

\begin{table}[t]
    \centering
    \renewcommand\arraystretch{1.5}
    \caption{The results of Top-1, Top-5 , Top-10, MAR and MFR of FL baselines and \appname.}
    \resizebox{1\columnwidth}{!}{ 
        \begin{tabular}{c|c|c|c|c|c|c|c}
        \hline
            Metrics &Scenario  & Ochiai & Dstar& Barinel& MLP-FL & CNN-FL & RNN-FL  \\
            \hline
            \multirow{2}{*}{Top-1} & baseline& 38& 38& 35 &9  &6&27\\           
            \multirow{2}{*}{} &\appname	&\pmb{50}	&\pmb{41}&\pmb{42} &\pmb{27 } &\pmb{34}&\pmb{34}\\
            \cline{1-8}
            
            \multirow{2}{*}{Top-5} & baseline& 113& 111& 110 & 51 &23&72\\         
            \multirow{2}{*}{} &\appname	&	\pmb{123}&\pmb{121}&\pmb{119}& \pmb{ 64}&\pmb{86}&\pmb{81}\\
            \cline{1-8}
            
            \multirow{2}{*}{Top-10} & baseline& 156&152 & 154 & 73 &29&98\\
            \multirow{2}{*}{} &\appname	&\pmb{177}&\pmb{160}&\pmb{167} & \pmb{86} &\pmb{116}&\pmb{111}\\
            \cline{1-8}
            
            \multirow{2}{*}{MFR} & baseline& 371.79& 363.80& 373.81 &  987.94&1514.86&602.75\\ 
            \multirow{2}{*}{} &\appname	&\pmb{227.70}&\pmb{261.58} &\pmb{211.70} & \pmb{259.24} &\pmb{479.01}&\pmb{296.04}\\
            \cline{1-8}
            
            \multirow{2}{*}{MAR} & baseline& 676.08& 704.25& 680.49 &   1319.10&   1823.73&	948.71 \\
            \multirow{2}{*}{} &\appname	&\pmb{307.82}&\pmb{338.76}&\pmb{283.07} & \pmb{369.70} &\pmb{549.95}&\pmb{333.76}\\
            \cline{1-8}
        \end{tabular}
    }
    \label{table2}
\end{table}
\subsection{Research Questions and Results}
To evaluate the effectiveness of our approach, we design and conduct the experiments to investigate the following research questions:
\begin{itemize}
    \item \textbf{RQ1:} How does \appname perform in localizing real faults compared with original state-of-the-art FL approaches? This RQ aims at investigating whether \appname improves FL effectiveness after applying our approach. If the effectiveness of the FL approach increases after applying \appname, it means that \appname can mitigate the effect of class imbalance in FL.
    \item \textbf{RQ2:} How effective is \appname as compared with the representative data optimization approaches? This RQ is to further verify the ability of \appname to mitigate the effect of class imbalance in FL via comparing other representative data optimization approaches. If \appname outperforms other representative data optimization approaches, it means that \appname is more effective than other representative approaches in addressing the class imbalance problem of FL.
    \item \textbf{RQ3:}  Does each component contributes to the effectiveness of \appname? This RQ is to check whether each component of \appname (\ie a GAN or a failure-inducing context) contributes to the effectiveness of \appname. We use three cases: original FL (denoted as baseline), \appname only using GAN (denoted as CGAN4FL(GAN)), \appname using GAN and failure-inducing context (denoted as CGAN4FL(GAN+context)). If we acquire a FL effectiveness relationship: CGAN4FL(GAN+context) > CGAN4FL(GAN) > baseline,
    it means that the GAN (due to CGAN4FL(GAN) > baseline) and the failure-inducing context (due to CGAN4FL(GAN+context) > CGAN4FL(GAN)) both contribute to \appname.
\end{itemize}
 
\textbf{RQ1. How does \appname perform in localizing real
faults compared with original state-of-the-art FL approaches?}

There are two main types of FL: spectrum-based fault
localization (SFL) and deep learning-based fault localization (DLFL). 
Recent studies~\cite{zhang2021study,pearson2017evaluating} have shown the most effective SFL approaches (\ie \textbf{Dstar}~\cite{wong2012software}, \textbf{Ochiai}~\cite{wong2016survey}, and \textbf{Barinel}~\cite{abreu2009spectrum}) and DLFL approaches (\ie \textbf{MLP-FL}~\cite{zheng2016fault}, \textbf{CNN-FL}~\cite{zhang2019cnn}, and \textbf{RNN-FL}~\cite{zhang2021study}) in locating real faults.
Thus, we use the six state-of-the-art FL approaches as the baselines and apply \appname to them to compare their effectiveness. 
For details of these FL approaches, please refer to Section~\ref{backgroundFL}.

\tablename~\ref{table2} shows the Top-K, MAR, and MFR results of the comparisons of the FL baselines and our approach \appname.
It illustrates two scenarios: a baseline without using \appname (referred to as baseline) and using \appname (referred to as \appname). 
For the convenience of reading, 
we bold the experimental results in the tables, indicating which approach performs better.

As shown in \tablename~\ref{table2},
\appname significantly outperforms all the baselines.  
For SFL approaches,
take Ochiai as an example.
The number of faults that \appname can locate is 50, 123, and 177 for the Top-1, Top-5, and Top-10 metrics, respectively.
The results denote the Top-1, Top-5, and Top-10 metrics have increased by 31.58$\%$, 8.85$\%$, and 13.46$\%$ as compared with Ochiai. 
For DLFL approaches,
take MLP-FL as an example. 
The number of faults that \appname can locate is 27, 64, and 86 for the Top-1, Top-5, and Top-10 metrics, respectively, 
\ie the Top-1, Top-5, and Top-10 metrics have increased by 200.00$\%$, 25.49$\%$, 17.81$\%$ as compared with the MLP-FL.
Furthermore,
the MFR and MAR metrics show that the rank of \appname is lower than that of baselines for all six FL techniques. The results show that \appname can always locate one buggy line first and find all buggy lines with the least effort.

\begin{figure}[htbp]    
\centerline{\includegraphics[scale=0.3]{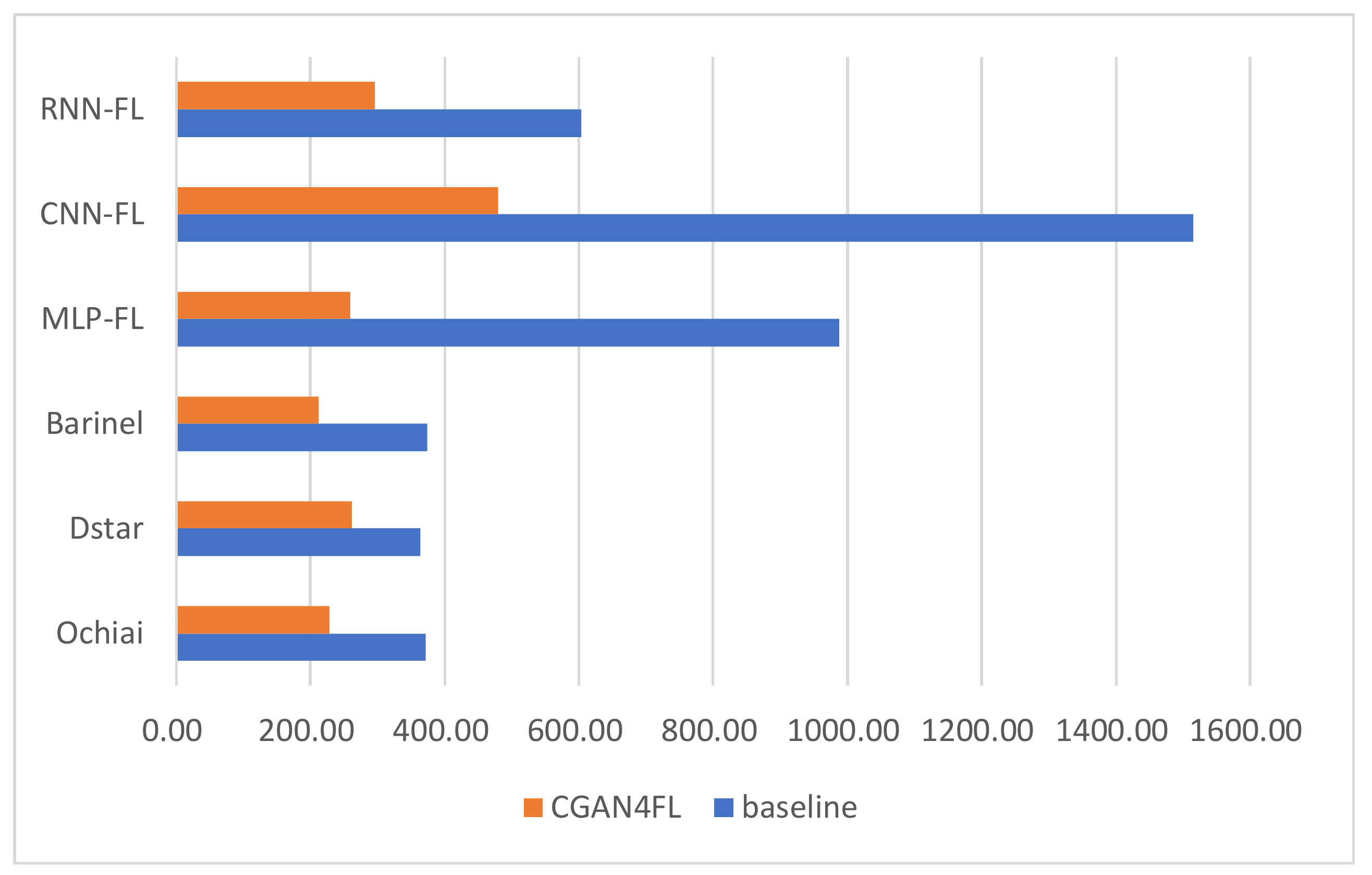}}
    \caption{MFR distribution of FL baselines and \appname}
    \label{fig6}
\end{figure}

\figurename~\ref{fig6} visually shows the MFR distribution of the FL baselines and \appname. The results show that \appname significantly improves FL effectiveness.

\begin{figure}[htbp]    
\centerline{\includegraphics[scale=0.3]{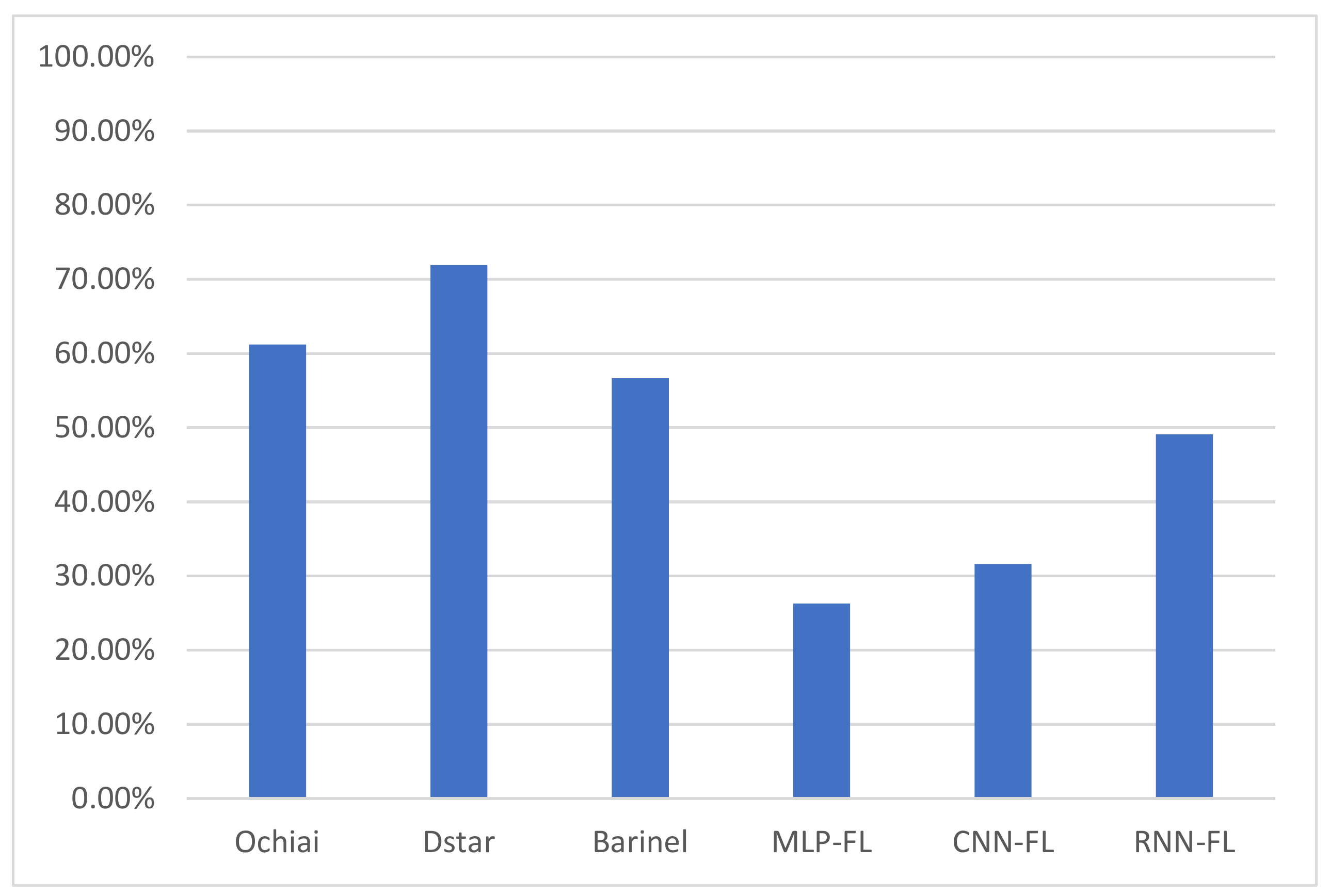}}
    \caption{RImp distribution of \appname vs FL baselines.}
    \label{fig7}
\end{figure}

\figurename~\ref{fig7} shows the RImp distribution of our approach \appname vs the six FL baselines. As shown in \figurename~\ref{fig7},
all RImp values are less than 100\%,
showing that our approach improves all the baselines after applying \appname.
Take the MLP-FL as an example. 
MLP-FL will examine 987.94 lines on average to locate the first bug in all faulty versions (\ie MFR), while \appname only checks 259.24 lines of code. 
Thus, the value of RImp is 26.24$\%$, indicating that for locating the first faulty statement, 
the number of statements to be checked by \appname is 26.24$\%$ of the original MLP-FL.
\begin{tcolorbox}[boxrule=0pt, frame empty]
 \emph{\textbf{Summary for RQ1}: In RQ1, we discuss the effectiveness of the six state-of-the-art FL approaches using \appname and without using \appname. 
 The experimental results show that 
\appname is effective to improve FL effectiveness by mitigating the effect of class imbalance in FL.}
\end{tcolorbox}
 
\textbf{RQ2. How effective is \appname as compared with the representative data optimization approaches?}

To further evaluate the ability of \appname to mitigate the effect of class imbalance in FL, we compare our approach with two representative data optimization approaches, \ie resampling~\cite{gao2013theoretical,zhang2017theoretical,zhang2021improving} and undersampling~\cite{wang2020ietcr}.
Resampling and undersampling acquire a class-balanced dataset by replicating minority samples and removing the majority samples, respectively. 
For more details, resampling  and undersampling can refer to Gao~\etal~ \cite{gao2013theoretical} and Wang~\etal~ \cite{wang2020ietcr}, respectively.

\tablename~\ref{table3} shows the Top-K, MAR, and MFR results of the two representative data optimization approaches and our approach \appname.
As shown in \tablename~\ref{table3}, \appname outperforms resampling
and undersampling in all cases of SFL and most cases of DLFL. 
Taking Ochiai as an example, the number of faults \appname can locate is 50, 123, and 177 for Top-1, Top-5, and Top-10 metrics, respectively. 
The results indicate that the Top-1, Top-5, and Top-10 metrics have increased by 284.62$\%$, 136.54$\%$, and 126.92$\%$ respectively as compared with undersampling, and increased by 47.06$\%$,33.70$\%$, and 40.48$\%$ respectively as compared with resampling. 
Furthermore,
the MFR and the MAR of \appname are lower than the two representative data optimization approaches in almost all cases, 
showing that \appname performs better than the two representative approaches. 

\begin{figure}[htbp]    
\centerline{\includegraphics[scale=0.3]{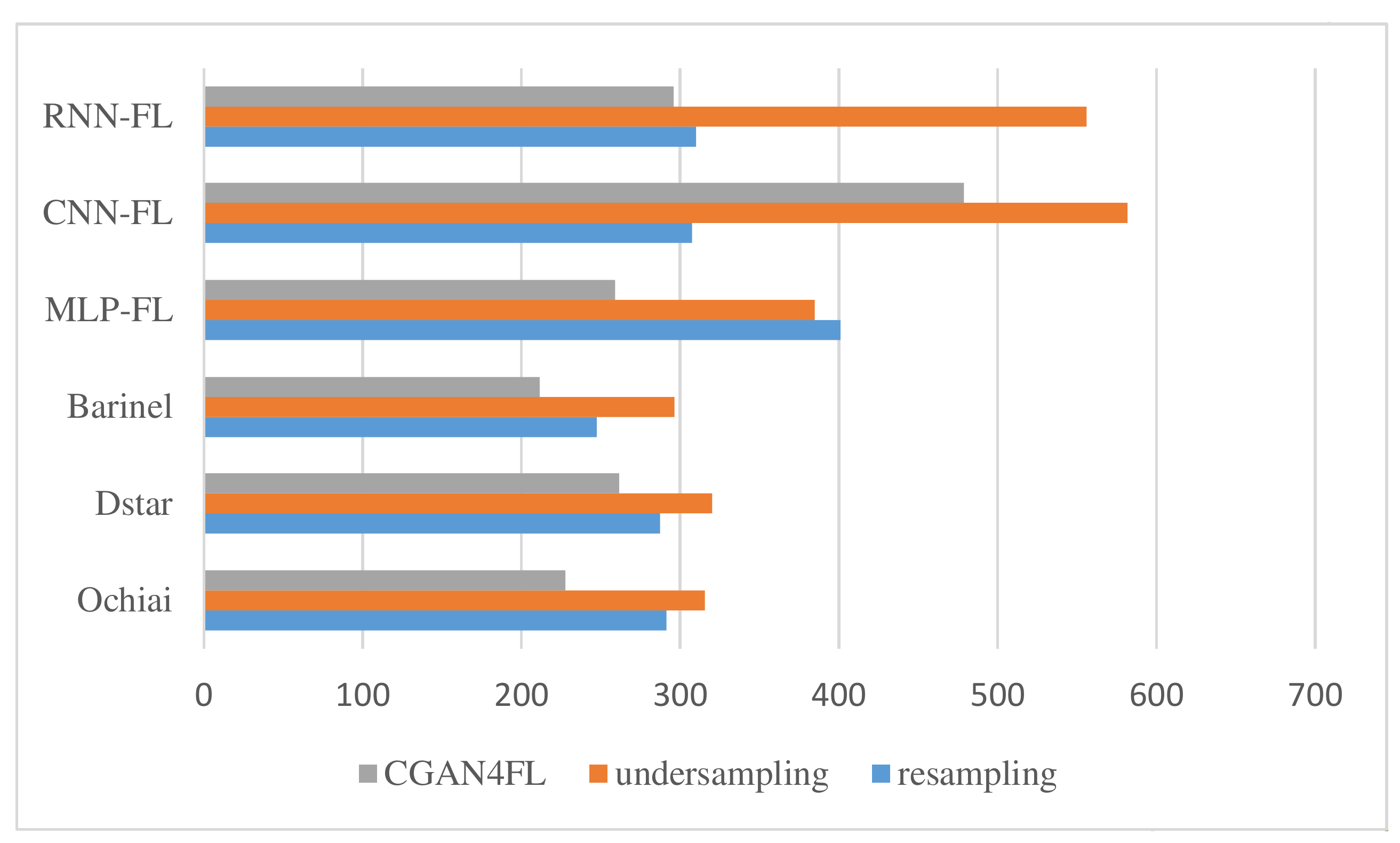}}
    \caption{MFR distribution of  resampling, undersampling and \appname}
    \label{fig8}
\end{figure}

\begin{figure}[htbp]    
\centerline{\includegraphics[scale=0.3]{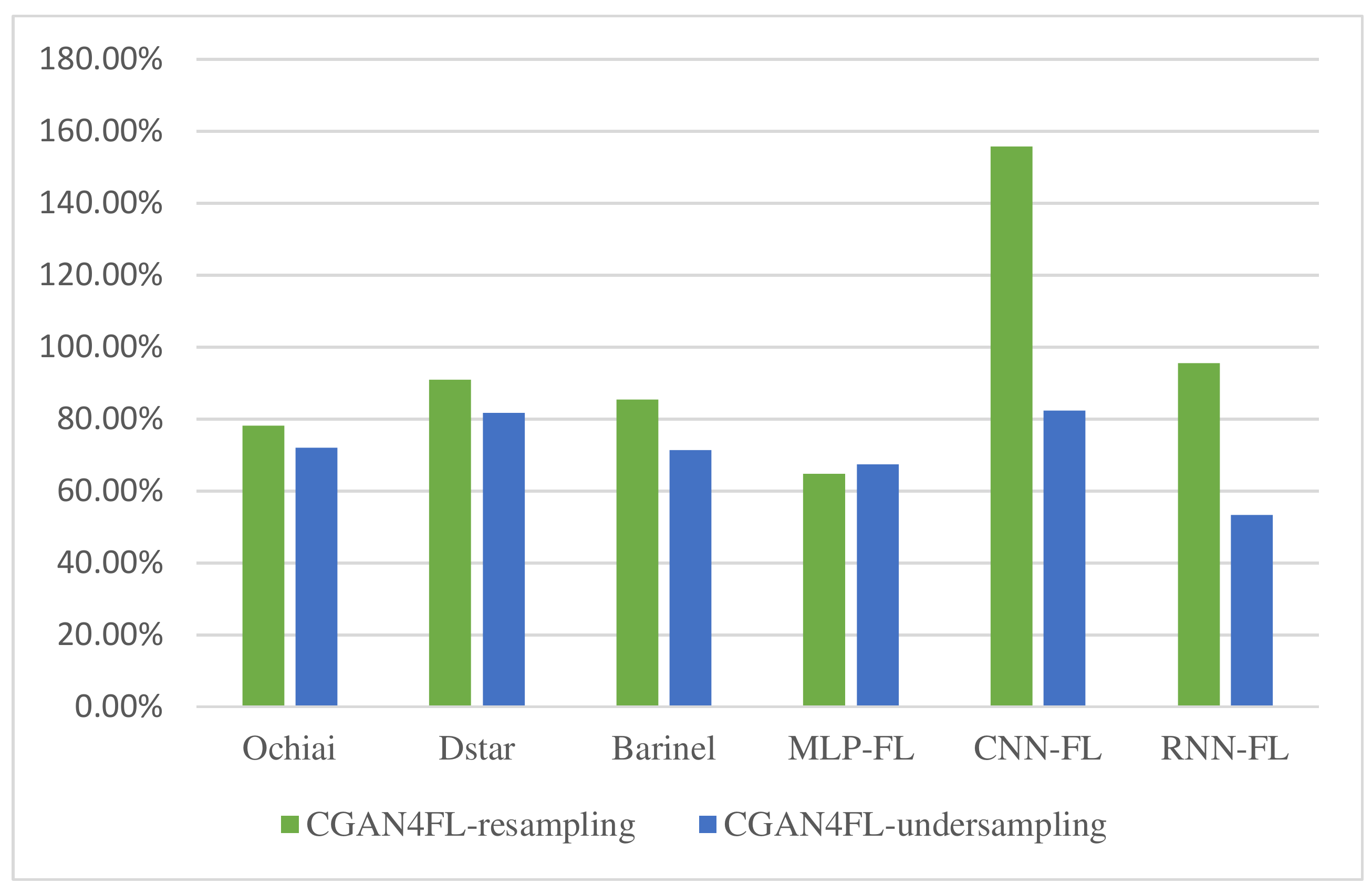}}
    \caption{RImp distribution of \appname vs resampling and \appname vs undersampling.}
    \label{fig9}
\end{figure}

Furthermore, \figurename~\ref{fig8} visually shows the MFR distribution of resampling, undersampling, and \appname,
indicating that \appname is the best of these three scenarios except for one case of resampling is better in CNN-FL. 
\figurename{}~\ref{fig9} shows the RImp distribution under two scenarios: \appname  vs resampling and \appname vs undersampling. 
As shown in \figurename{}~\ref{fig9},  the RImp values of all the cases are less than 100\% except for one case of \appname vs resampling in CNN-FL. 
Thus, 
we can conclude that \appname performs better than resampling
and undersampling in addressing the class imbalance problem in FL.
\begin{tcolorbox}[boxrule=0pt, frame empty]
 \emph{\textbf{Summary for RQ2}: In RQ2, we compare \appname with two representative data optimization approaches, \ie resampling and undersampling. The experimental results show that \appname is more effective than the two representative data optimization approaches in almost all cases for mitigating the effect of class imbalance in FL.}
\end{tcolorbox}

\begin{table}[t]
    \centering
    \renewcommand\arraystretch{1.5}
    \caption{The results of TOP-1, TOP-5, TOP-10, MAR and MFR of data optimization approaches and \appname.}
    \resizebox{1\columnwidth}{!}{ 
        \begin{tabular}{c|c|c|c|c|c|c|c}
        \hline
            Metrics &Scenario  & Ochiai & Dstar& Barinel& MLP-FL & CNN-FL & RNN-FL  \\
            \hline
            \multirow{3}{*}{Top-1} & resampling& 34& 34& 33 & 8 &\pmb{34}&\pmb{34}\\
            \multirow{3}{*}{} &undersampling	&13&13&13 &15  &	9 &	14\\
            \multirow{3}{*}{} &\appname	&\pmb{50}	&\pmb{41}&\pmb{42} & \pmb{27} &\pmb{34}&\pmb{34}\\
            \cline{1-8}

            \multirow{3}{*}{Top-5} & resampling& 92& 92& 97 & 53 &\pmb{92}&\pmb{92}\\
            \multirow{3}{*}{} &undersampling	&52&52&51 &50  &	37 &51\\
            \multirow{3}{*}{} &\appname	&\pmb{123}&\pmb{121}&\pmb{119}& \pmb{64} &86&81\\
            \cline{1-8}
            
            \multirow{3}{*}{Top-10} & resampling& 126& 126& 139 & 73 &\pmb{125}&\pmb{128}\\
            
            \multirow{3}{*}{} &undersampling	&78&78&79 &69  &	58 &	80\\
            
            \multirow{3}{*}{} &\appname	&\pmb{177}&\pmb{160}&\pmb{167} &\pmb{86}  &116&111\\
            \cline{1-8}
            
            \multirow{3}{*}{MFR} & resampling& 291.58	&287.53&247.59&	400.81&	\pmb{307.36}&310.00\\
            
            \multirow{3}{*}{} &undersampling	&315.66&320.02&296.23 &384.75  &582.04 &	555.91 \\
            
            \multirow{3}{*}{} &\appname	&\pmb{227.70}&\pmb{261.58}&\pmb{211.70 }& \pmb{259.24} &479.01&\pmb{296.04}\\
            \cline{1-8}
            
            \multirow{3}{*}{MAR} & resampling& 625.38& 668.06& 575.41 &   736.89&   660.98&	685.14 \\
            
            \multirow{3}{*}{} &undersampling	& 711.35&732.99 & 680.32 &   774.98&   1014.07&	971.23 \\
            
            \multirow{3}{*}{} &\appname	&\pmb{307.82}&\pmb{338.76}&\pmb{283.07} & \pmb{369.70 }&\pmb{549.95}&\pmb{333.76}\\
            \cline{1-8}
        \end{tabular}
    }
    
    \label{table3}
\end{table}
\textbf{RQ3. Does each component contributes to the effectiveness of \appname?}

To check whether each component contributes to the effectiveness of \appname, we use Wilcoxon-Signed-Rank Test (WSR)~\cite{wilcoxon1992individual} to verify whether the effectiveness relationship (\ie CGAN4FL(GAN+context) > CGAN4FL(GAN) > baseline) is satisfied or not,
where CGAN4FL(GAN+context), CGAN4FL(GAN), and baseline denote \appname using GAN and failure-inducing context, \appname only using GAN, and original FL respectively.
For each FL approach, 
we perform two paired Wilcoxon-Signed-Rank tests (\ie CGAN4FL(GAN+context) vs CGAN4FL(GAN) and CGAN4FL(GAN) vs baseline) by using the ranks of the faulty statements
as the pairs of measurements.

\tablename~\ref{table4} shows the statistical results of all the tests at the  $\sigma$ level of 0.05.
The `conclusion’ column gives the conclusion according to $p$-value. 
Take Ochiai as an example.
For CGAN4FL(GAN) vs baseline, 
the the $p$-value of greater, less, and two-sided are 1, 2.68e-08, and 5.36e-08 respectively. According to the definition of WSR, it means that the MFR value of CGAN4FL(GAN) (\ie Ochiai using CGAN4FL(GAN)) is less than that of the baseline (\ie original Ochiai), leading to a BETTER result.
For CGAN4FL(GAN+context) vs CGAN4FL(GAN), the $p$-value of greater, less, and two-sided are 1, 1.43e-03, and 2.86e-03  respectively.  It means that the MFR value of CGAN4FL(GAN+context) (\ie Ochiai using our approach \appname) is less than that of CGAN4FL(GAN) (\ie Ochiai using CGAN4FL(GAN)), also leading to a BETTER result. 
From the table, we can observe that both CGAN4FL(GAN) vs baseline and CGAN4FL(GAN+context) vs \appname(GAN) obtain BETTER results in all cases, 
meaning that the effectiveness relationship CGAN4FL(GAN+context) > CGAN4FL(GAN) > baseline is satisfied.
Thus, each component of \appname contributes to its effectiveness.

\begin{tcolorbox}[boxrule=0pt, frame empty]
 \emph{\textbf{Summary for RQ3}: In RQ3, we make statistical comparisons in two scenarios: CGAN4FL(GAN+context) vs CGAN4FL(GAN) and CGAN4FL(GAN) vs baseline. 
 The experimental results show that the effectiveness relationship CGAN4FL(GAN+context) > CGAN4FL(GAN) > baseline holds. Thus, we can conclude that each component of \appname (\ie failure-inducing context and GAN) contributes to its effectiveness.}
\end{tcolorbox}
\begin{table}[]
\centering
    \renewcommand\arraystretch{1.5}
    \caption{Statistical comparison of CGAN4FL(GAN) vs baseline and CGAN4FL(GAN+context) vs CGAN4FL(GAN).}
    \resizebox{1\columnwidth}{!}{ 
        \begin{tabular}{c|c|cccc}

        \hline
        method                 &comparison & greater & less & two-sided   & conclusion       \\ \hline
        \multirow{2}{*}{Ochiai}   &{\appname}(GAN) vs baseline    &1 & 2.68e-08	&5.36e-08	 &\textbf{BETTER} \\ 
                                 & {\appname}(GAN+context) vs {\appname}(GAN)        &1 & 1.43e-03	&2.86e-03 &\textbf{BETTER}\\ \hline
                                 
        \multirow{2}{*}{Dstar}  & {\appname}(GAN) vs baseline   & 1 & 6.07e-06	& 1.21e-05& \textbf{BETTER}\\ 
                                 &  {\appname}(GAN+context) vs {\appname}(GAN)    &1 & 4.14e-10	 &8.28e-10	 &\textbf{BETTER}\\ \hline

        \multirow{2}{*}{Barinel}    & {\appname}(GAN) vs baseline    &1 &8.05e-04	&1.61e-03  &\textbf{BETTER} \\ 
                                 & {\appname}(GAN+context) vs {\appname}(GAN)       &1 &3.57e-03	 & 7.13e-03&\textbf{BETTER} \\ \hline
        \multirow{2}{*}{MLP-FL}    & {\appname}(GAN) vs baseline    &1 &1.42e-02	 &2.84e-02	 &\textbf{BETTER} \\ 
                                 &{\appname}(GAN+context) vs {\appname}(GAN)        &1 &1.84e-10	& 3.67e-10 &\textbf{BETTER}\\ \hline
        \multirow{2}{*}{CNN-FL} & {\appname}(GAN) vs baseline   &1 & 1.70e-07	& 3.39e-07&\textbf{BETTER}\\ 
                                 & {\appname}(GAN+context) vs {\appname}(GAN)          &1 &1.35e-05	 &2.71e-05 & \textbf{BETTER}\\ \hline
        \multirow{2}{*}{RNN-FL}    & {\appname}(GAN) vs baseline   &1 &9.87e-04	 & 1.97e-03& \textbf{BETTER}\\ 
                                 & {\appname}(GAN+context) vs {\appname}(GAN)          &1 & 3.07e-13	& 6.14e-13&\textbf{BETTER} \\ \hline
        
        \end{tabular}
    }
    \label{table4}
\end{table}
\section{Discussion}\label{discussion}
\subsection{Threats to Validity}

\textbf{The implementation of baselines and our approach.} Our implementation of baselines and \appname may potentially contain bugs. 
For the three SFL approaches (\ie Dstar, Ochiai, and Barinel), 
we implement them according to their formulas and then manually test their correctness. 
For the three DLFL approaches (\ie  MLP-FL, RNN-FL, and CNN-FL),
we acquire the source code of CNN-FL from the authors and implement the other two DLFL approaches via replacing the deep learning component with MLP and RNN from the source code of CNN-FL. 
Since a neural network has many parameters for its construction (\eg learning rate, batch size), 
some parameters of MLP-FL, RNN-FL, and CNN-FL may differ from the original paper. 
Besides the implementation of baselines, 
we implement our pipeline of failure-inducing context construction, context-aware GAN model training, and class-balanced raw data generation, which may also potentially include bugs. 
To mitigate those threats, we check our code implementation rigorously and make all relevant code publicly available (see the footnote in Section~\ref{intro}).

\textbf{The generalizability.} We conduct our experiments on the real faults dataset benchmark, Defects4J, which is widely used in fault localization and program repair community.  
Although the subject programs selected in our experiments are all from the real world and our approach performs well on these programs, 
it may be not effective for other programs since no dataset can cover all possible cases of faults in practice. 
Thus,
it is worthwhile to conduct more experiments on more large-sized programs with real faults to further verify the effectiveness of our approach in mitigating the effect of class imbalance in FL.

\subsection{Reasons for \appname Is Effective}
Our experiments demonstrate that \appname is more effective than the compared baselines.
The main reasons are threefold:
(1) \appname takes full advantage of the program dependency to capture the statements closely related to the faulty statements.
In other words, the utilization of program slicing removes the fault-irrelevant statements precisely.
(2) The GAN is a powerful model that could generate synthesized samples that are like real samples.
More importantly, we provide the GAN model with the expert knowledge in software debugging (\ie the fault-relevant statements obtained by program slicing), which could potentially improve the accuracy and efficiency of the model.
(3) The advantages of both the program analysis technique (\ie the program slicing) and the advanced generative network (\ie the GAN model) are well combined by \appname to gain the high-quality generated input data.
Further, the high-quality and class-balanced input data could be beneficial to the state-of-the-art SFL and DLFL approaches.


\section{Related work}\label{relatedwork}
\subsection{Fault Localization}
Spectrum-based fault localization (SFL)~\cite{naish2011model,abreu2007accuracy} and Deep learning-based fault localization (DLFL)~\cite{zhang2019cnn,zhang2021study,lou2021boosting} are the two most popular FL approaches.
Researchers have proposed many SFL techniques (\eg  Tarantula\cite{jones2005empirical}, Ochiai\cite{abreu2007accuracy}, Jaccard\cite{abreu2007accuracy}, and DStar\cite{wong2013dstar}),
and enhanced program spectrum (\eg \cite{dallmeier2005lightweight,yilmaz2008time}) and a test suite (\eg \cite{li2019deepfl,wong2009bp}) for more improvement.
With a large number of SFL techniques, 
many studies~\cite{abreu2007accuracy,le2013theory,naish2011model,wong2013dstar} have explored the best SFL techniques. 
Yoo \etal~\cite{yoo2014no} have found that 
there is no SFL technique claiming that it can outperform all others under every scenario.
Even if the best SFL technique does not exist.
The existing studies~\cite{xie2013theoretical,Xie2013Provably} have found a group of optimal SFL techniques, \ie the group of optimal SFL techniques cannot outperform each other whereas they can outperform all the other SFL techniques outside the group.
DLFL~\cite{zhang2019cnn,zhang2021study,lou2021boosting} uses deep learning to locate a fault and recently attracts much attention. Wang~\etal propose the FL approach BPNN-FL\cite{wong2007effective} using BP neural network model as a pipeline for learning input and output relationships. 
Then, Wong~\etal~\cite{wong2009bp} improve their BPNN-FL approach by removing irrelevant statements. 
Similar to the idea of Wong~\etal, 
many researchers directly use the raw data as training data, 
and propose different DLFL approaches using different neural networks (\eg MLP-FL\cite{zheng2016fault}, CNN-FL\cite{zhang2019cnn}, and RNN-FL\cite{zhang2021study}).
In contrast to devising an effective FL approach,
our work focuses on addressing the class imbalance problem in FL and can be used in tandem with these FL approaches. 

\subsection{Class Imbalance}
In recent decades, 
researchers have proposed many methods to address the class imbalance problem. 
The typical methods are data-level, algorithm-level, hybrid, and ensemble learning methods. 
The data-level methods add a preprocessing step to mitigate the effect of class imbalance in the learning process\cite{batista2004study,lee2016plankton,buda2018systematic}. 
The algorithm-level methods create or modify deep learning algorithms for addressing the class imbalance problem\cite{wang2016training,lin2017focal}. 
The hybrid methods combine algorithm-level and data-level methods~\cite{huang2016learning,ando2017deep,dong2018imbalanced}.
The ensemble learning methods~\cite{polikar2006ensemble,rokach2010ensemble} use ensembles to increase the accuracy of classification by training several different classifiers and combining their decisions to output a single class label.
Specifically, with regard to ensemble learning methods,
there are lots of different approaches, \eg SMOTEBoost\cite{chawla2003smoteboost}, RUSBoost\cite{seiffert2009rusboost}, IIVotes\cite{blaszczynski2010integrating}, EasyEnsemble\cite{liu2008exploratory}, and SMOTEBagging\cite{wang2009diversity}.
These works focus on addressing the class imbalance problem in the artificial intelligence field.
In contrast,
our work focuses on mitigating the effect of class imbalance in a different research field (\ie fault localization).

\section{Conclusion}\label{conclusion}
In this paper, we propose \appname: a data augmentation approach that uses context-aware GAN to mitigate the effect of the class imbalance in FL. 
\appname embraces two main ideas: 
(1) data augmentation is a potential and effective solution to the class imbalance problem in FL;
(2) a failure-inducing context is useful for guiding and acquiring a more precise data augmentation process.
To implement the above ideas,
we use program dependencies to construct a failure-inducing context showing how a failure is caused, and integrate the context into a generative adversarial network to learn the features of minority class and synthesize minority class data for generating a class-balanced dataset for FL.
The experiments show that our approach \appname  is effective to mitigate the effect of class imbalance in FL.

In the future, we intend to use more large-sized program to further verify our approach, and explore other generative networks for more improvement.
\section*{Acknowledgment}
This work is partially supported by the National Natural Science Foundation of China (No. 62272072), the Fundamental
Research Funds for the Central Universities (No. 2022CDJDX-005), and the Major Key Project of PCL (No. PCL2021A06).

\bibliographystyle{IEEEtran}
\bibliography{ref}
\end{document}